# Diffusion geometry approach to efficiently remove electrical stimulation artifacts in intracranial electroencephalography

Sankaraleengam Alagapan[1,2], Hae Won Shin [3,4], Flavio Fröhlich [1,2,3,5,6,7], Hau-tieng Wu [8,9] *

Correspondence should be addressed to: Hau-tieng Wu, 140 Science Drive, Physics building, Room 207 Durham, NC, 27708. Email: hauwu@math.duke.edu

1    Carolina Center for Neurostimulation, University of North Carolina at Chapel Hill, Chapel Hill, NC, USA

2    Department of Psychiatry, University of North Carolina at Chapel Hill, Chapel Hill, NC, USA

3    Department of Neurology, University of North Carolina at Chapel Hill, Chapel Hill, NC, USA

4    Department of Neurology, University of North Carolina at Chapel Hill, Chapel Hill, NC, USA

5    Department of Cell Biology and Physiology, University of North Carolina at Chapel Hill, Chapel Hill, NC, USA

6    Department of Biomedical Engineering, University of North Carolina at Chapel Hill, Chapel Hill, NC, USA

7    Neuroscience Center, University of North Carolina at Chapel Hill, Chapel Hill, NC, USA

8    Department of Mathematics, Duke University, Durham, NC, USA

9    Department of Statistical Science, Duke University, Durham, NC, USA

# Abstract

**Objective:** Cortical oscillations, electrophysiological activity patterns, associated with cognitive functions and impaired in many psychiatric disorders can be observed in intracranial electroencephalography (iEEG). Direct cortical stimulation (DCS) may directly target these oscillations and may serve as therapeutic approaches to restore functional impairments. However, the presence of electrical stimulation artifacts in neurophysiological data limits the analysis of the effects of stimulation. Currently available methods suffer in performance in the presence of nonstationarity inherent in biological data.

**Approach:** Our algorithm, Shape Adaptive Nonlocal Artifact Removal (SANAR) is based on unsupervised manifold learning. By estimating the Euclidean median of k-nearest neighbors of each artifact in a nonlocal fashion, we obtain a faithful representation of the artifact which is then subtracted. This approach overcomes the challenges presented by nonstationarity.

**Main results:** SANAR is effective in removing stimulation artifacts in the time domain while preserving the spectral content of the endogenous neurophysiological signal. We demonstrate the performance in a simulated dataset as well as in human iEEG data. Using two quantitative measures, that capture how much of information from endogenous activity is retained, we demonstrate that SANAR's performance exceeds that of one of the widely used approaches, independent component analysis, in the time domain as well as the frequency domain.

**Significance:** This approach allows for the analysis of iEEG data, single channel or multiple channels, during DCS, a crucial step in advancing our understanding of the effects of periodic stimulation and developing new therapies.

# Introduction

Electrical stimulation has become an important tool for localization of human brain function. Noninvasive stimulation methods like transcranial magnetic stimulation (TMS) and transcranial alternating current stimulation (tACS) have enabled modulation of large-scale network dynamics to target behavior (Frohlich, 2014; Luber and Lisanby, 2014). In addition, deep brain stimulation (DBS) and direct cortical stimulation (DCS), both invasive methods, have been used for treatment of neurological and psychiatric disorders (Kocabicak et al., 2015), mapping of cortical function (Borchers et al., 2012; Duffau et al., 1999; Matsumoto et al., 2004) as well as modulation of memory (Ezzyat et al., 2017; Ezzyat et al., 2018; Kucewicz et al., 2018). Typically, these studies have investigated the effect of stimulation on behavior, where there is limited insight into the effect of stimulation on ongoing brain activity. The effect of stimulation is generally measured using electroencephalography (EEG), magnetoencephalography (MEG) or invasive electroencephalography (iEEG). However, the presence of high amplitude stimulation artifacts caused by the interaction between the electric field and the recording system mask the endogenous activity during stimulation. Stimulation artifacts are reflected in the frequency domain, and a structured spiky pattern appears when the stimulation follows a periodic structure. Since most of the analyses of brain activity occur in frequency domain, the presence of artifacts renders the analysis of stimulation effects on oscillations useless. Thus, removal of these artifacts would enable us gain insight into the interaction between electrical stimulation and neural activity and in turn, neural activity and behavior.

There have been several proposals to remove these stimulation artifacts post-hoc, for example, the template subtraction (TS) algorithm (Alagapan et al., 2016; Hashimoto et al., 2002; Qian et al., 2017; Trebaul et al., 2016; Wichmann, 2000), the principal component analysis (PCA) (Helfrich et al., 2014; ter Braack et al., 2013) or the independent component analysis (ICA)(Albouy et al., 2017; Lu et al., 2012; Rogasch et al., 2014) in the case of multi-channel recordings. Apart from these approaches, Kalman Filtering have been used to suppress stimulation artifacts in neurophysiological data (Morbidi et al., 2007; Morbidi et al., 2008). The approach involves fitting separate generative models for the artifact and for the neurophysiological data and applying the Kalman filter to extract the artifact-free data of interest. Yet another approach is to use the spectral information of the artifacts using matched filters (Allen et al., 2010; Sun et al., 2014) or empirical mode decomposition (Al-ani et al., 2011; Santillan-Guzman et al., 2013) to separate neurophysiological signal from artifacts.

While these methods have been successfully applied, there are some limitations. The TS algorithm tends to suffer from the template bias, which arises from the possible deviation of the designed template from the ground truth. Due to the non-stationarity of the physiological system, the artifact might vary from one to the other, and it might not be possible to find a universal template. PCA and ICA algorithms tend to produce poorer results when the recording gets longer, as the non-stationarity of the endogenous activity tends to increase in longer recordings. This limitation is a direct consequence of the underlying stationarity assumption of the techniques. While it is possible to truncate signals into pieces and process each piece separately, how to "glue" together all pieces could be another challenge. These limitations in general downgrade the quality of the recovered neurophysiological signal. Note that while the TS algorithm could be applied to a single channel signal, ICA and PCA based algorithms need multiple channels. The Kalman filter approach is limited by the model that is fit for the artifact and is also susceptible to nonstationarity in artifact shapes.

To overcome these limitations, which are inherited from the non-stationarity nature of the physiological system, we propose a novel artifact removal algorithm, the Shape Adaptive Nonlocal Artifact Removal (SANAR), based on the manifold model commonly used in the machine learning field. Briefly, to fully capture the artifact behavior, we acknowledge that while the artifacts look similar, they exhibit variations across time and trials due to the non-stationarity. We therefore capture the variation among artifacts by a low dimensional and nonlinear geometric model. Based on this model, the algorithm recovers the artifact by respecting this nonlinear structure; that is, the artifact is recovered by taking the median of similar artifacts parametrized by the manifold. On a high level, this algorithm could be understood as a variation of the TS algorithm, while we design a good "metric" to determine the template in an adaptive fashion. Indeed, for each artifact, we construct an exclusive template from those artifacts that are similar to the given artifact determined by the designed metric. By following this estimation of artifact with a simple linear removal, the endogenous neural activity is recovered.

In this paper, we provide details of SANAR, a brief mathematical basis and a demonstration of the algorithm in a simulation. We demonstrate SANAR applied to artifacts produced by DCS in iEEG and compare it with another approach based on ICA. While there are additional approaches, we focused on ICA is it is currently the most widely adopted approach. In addition, we provide two measures that capture the efficacy of SANAR in suppressing the artifacts and use them to quantify the performance difference between our approach and the ICA-based approach. To the best of our knowledge, this kind of performance measurement is seldom considered in the field (But see (Korhonen et al., 2011)).

## Materials and Methods

**Direct Cortical Stimulation and iEEG.** All experimental procedures were approved by the Institutional Review Board of University of North Carolina at Chapel Hill (IRB Number 13-2710). iEEG was recorded from 114 electrodes implanted in a participant performing a working memory task while being simultaneously stimulated. iEEG was recorded using a high-density EEG system (NetAmps 410, Electrical Geodesics Inc, Eugene, Oregon, United States). Sampling rate was set at 1000 Hz. The amplifier has a software anti-aliasing filter with a cutoff frequency at 500 Hz in addition to the inbuilt hardware anti-aliasing filter with cutoff at 4000 Hz. Electrical stimulation was applied between pairs of adjacent recording electrodes and consisted of 5-second-long pulse trains at with 110 ms between pulses (~9.1 Hz). Each biphasic pulse was 2 mA in amplitude and 400 μs in duration. The pulse trains were generated by a cortical stimulator (Cerestim M96, Blackrock Microsystems, Salt Lake City, Utah, United States). A total of 2 different electrode pairs were stimulated and each electrode pair was stimulated 20 times. The location of the recording electrodes and stimulation electrodes are shown in Figure 1A.

The stimulation produces transient artifacts that vary in amplitude i.e., as the distance from the stimulating electrodes increases, the stimulation artifact amplitude decreases (Figure 1B). In most practical situations, it is impossible to recover any physiological data from the stimulation electrodes (not shown in Fig 1B) and hence, we restrict our analyses to non-stimulation electrodes. Figure 1C provides examples of traces from electrodes that exhibit different artifact amplitudes relative to endogenous activity amplitude.

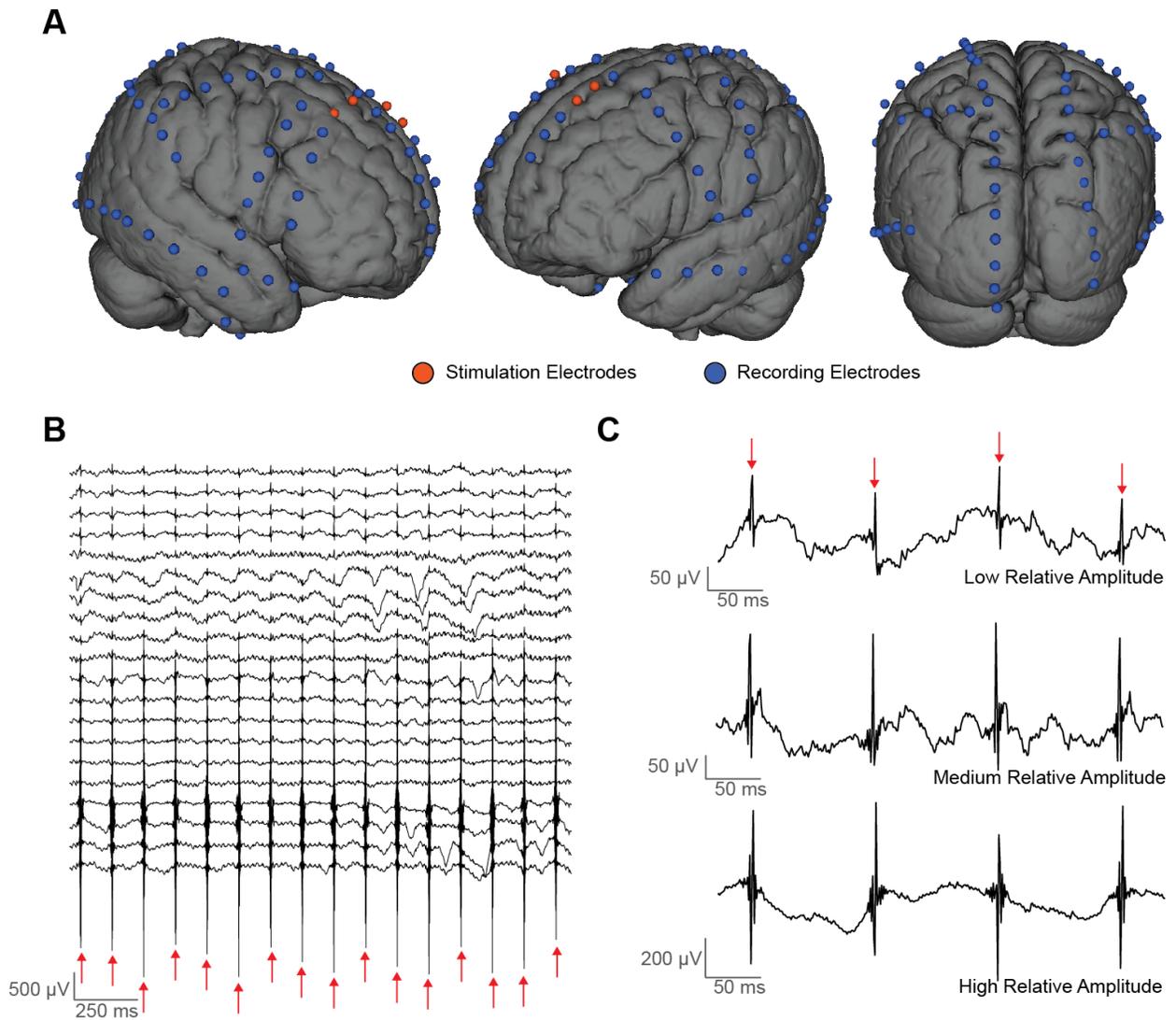

**Figure 1.** (A) Schematic of the location of electrodes on participant's brain. The orange electrodes denote stimulation electrodes while the blue electrodes denote recording electrodes. (B) Multichannel waveform of iEEG showing the differences in amplitudes of stimulation artifacts (red arrows) (C) Example traces of artifacts exhibiting different amplitudes relative to endogenous activity.

**Simulated EEG signal and stimulation.** We used a 'phantom' setup to simulate iEEG and DCS where the 'endogenous' signal is already known (Figure 3A). We used a saline solution (0.15 M KCL) to simulate the conductivity of the gray matter and placed an antenna connected to a function generator (SKMI, Taiwan) in the saline solution to act as a virtual dipole. A sine wave was generated by the function generator and this served as the ground truth signal. We immersed a strip consisting of 4 electrodes that is used for ECoG in the saline. The electrodes were connected to the stimulator and amplifier setup used in our experiments. Stimulation was applied through one pair of electrodes while the other two electrodes served as recording electrodes after which the stimulation electrodes and recording electrodes were swapped functionally. The sine wave frequency $F_{signal}$ was set at 7 Hz and 5 Hz. Stimulation frequency $F_{stim}$ was set at 5 Hz and 10 Hz and duration was set at 40 seconds. A total of 6 trials were collected ($F_{signal}$ 7 Hz, $F_{stim}$ 10 Hz; $F_{signal}$ 7 Hz, $F_{stim}$ 5 Hz; $F_{signal}$ 5 Hz, $F_{stim}$ 10 Hz) resulting in 12 traces for testing the algorithm.

**Computation Setup.** All computations for the presented work including algorithm development, analysis and statistics were carried out using Matlab (Mathworks Inc, Natick, MA, USA). A desktop computer with quad core processor (Intel Core i7-4770k CPU) and 32 GB memory running Windows 7 Enterprise was used for all analysis.

**Line noise removal using curve fitting.** We adopted a curve fitting approach to remove line noise (60 Hz) from the recording. This step preceded the stimulation artifact removal. This approach is advantageous compared to notch filtering since notch filtering may introduce distortion in artifact waveform (Luo and Johnston, 2010; Mitra and Pesaran, 1999) and in the context of our algorithm, yielded better performance (Figure S3). We fit a sine wave with 60 Hz as frequency to the iEEG data from each electrode and each trial separately. A least square cost-function was used to estimate the amplitude and phase offset of the sine wave. Then the fitted sine wave was subtracted from the raw data to remove the line noise.

**Shape Adaptive Nonlocal Artifact Removal.** The proposed SANAR algorithm removes the artifact incurred from DCS by combining the manifold model and the nonlocal Euclidean median algorithm (Chaudhury and Singer, 2012). The basic idea is similar to the template subtraction (TS) algorithm – find a template for the artifact and recover the EEG signal by subtracting the template from the recorded iEEG signal. However, in the proposed artifact removal algorithm, we account for the structure hidden inside the artifact pattern - different artifact waveforms are not linearly related and thus cannot be represented by a unique template with linearly transformation. Based on this structure, we design an exclusive template for each artifact by designing a metric (See Figure 2 for an illustration).

Suppose the stimulation happens at times $t_1 < t_2 < \cdots < t_n$, and we assume that $t_i - t_{i-1}$ is sufficiently large so that two consecutive stimulation are separated far apart. The periodic stimulation experiment carried out in this study fulfills this criterion. In our setup, while we know when each pulse train (a set of 10 or 20 pulses) starts and how long it is but we do not know the exact timing of each pulse. We can guess based on the start time and duration, but it is not perfect, and we need the help of a peak detection algorithm. Inspired by the TS idea, we divide the recorded iEEG signal, $X(t)$, sampled at 1000Hz, into non-overlapping segments, so that each segment contains one artifact. Note that over each segment, $I_i \subset \mathbb{R}$, the iEEG is composed of at least three components – the unwanted artifact $A_i$, the wanted iEEG signal $E_i$, and the inevitable noise $N_i$. In other words, if $X_i$ is the restriction of $X(t)$ on $I_i$, then $X_i = A_i + E_i + N_i$. Note that we do not assume what the artifact looks like; it can be of different pattern, as long as we can isolate each artifact. The basic idea of the proposed algorithm is composed of two steps. First, for each $X_i$, find $X_j$ so that $A_j$ is the same or similar to $A_i$. This step is based on designing a metric that is not sensitive to the existence of $E_i + N_i$. Note that the existence of $E_i + N_i$ is the main difficulty of designing this metric. In this work, we apply the modern machine learning techniques, including optimal shrinkage (Gavish and Donoho, 2017) and diffusion distance (Coifman and Lafon, 2006; El Karoui and Wu, 2016; Singer and Wu, 2017) to remove $E_i + N_i$ from $X_j$ so that we can faithfully determine $A_j$ that is the same or similar to $A_i$. Second, we find $X_{i_j}$, $j = 1, \ldots, K$, so that $A_{i_j}$ is the same or similar to $A_i$, then recover $A_i$ from $X_i$ by evaluating the median of $X_{i_j}$, $j = 1, \ldots, K$, at each sampling point. The step-by-step algorithm is detailed below.

1. Preprocessing the iEEG signal by removing the trend. The trend is estimated by the median filter with a window of length 200 ms, followed by a window smoothing of length 10 msec. To better align the stimulation artifacts when the artifact pattern is spiky, the signal is upsampled to 8,000 Hz (Laguna and Sornmo, 2000). To implement the upsampling by the ratio of p/q, where p and q are coprime integer numbers and p>q, the input data is first upsampled by a factor of the integer p by inserting zeros. Then a least-squares linear-phase FIR filter designed with a Kaiser window, where the parameter for the window shape is set to 5, is applied to the upsampled data. Finally, the result is downsampled by a factor of the integer q by throwing away samples.

2. Divide the recorded iEEG signal into segments $X_i$ according to the stimulation pattern, where $i = 1, \ldots, n$, and $n$ is the number of all artifact cycles. Specifically, the following two sub-steps are carried out. First, if the timestamps of stimulation artifacts are unknown to us, detect the peaks of the stimulation artifacts using a peak detection algorithm in the electrode that has the strongest artifacts. If there are multiple channels, since the artifacts occur concurrently across all the electrodes, we use the same locations for all electrodes. Denote the timestamp of the $i$-th stimulation artifact as $t_i$. Second, take $L = round[median(\{t_i - t_{i-1}\}_{i=2}^n)/8]$ and set $X_i$ to be the signal over the period $[t_i - L, t_i + L]$. Here $X_i$ can come from different channels, if there is more than one channel. Form a $p$ by $n$ data matrix $X$, where $p = 2L + 1$ is the length of each segment and the $i$-th column is the $i$-th artifact cycle.

3. Apply the singular value optimal shrinkage (Gavish and Donoho, 2017) filter on $X$ to remove the randomness from each $X_i$. Specifically, there are four sub-steps. First, estimate the standard deviation of the iEEG signal over the interval without stimulation artifact contamination; that is, take the iEEG signal over $[t_i + L, t_{i+1} - L]$ and evaluate its standard deviation $\sigma$. Second, run the singular value decomposition on $X/\sigma\sqrt{n}$ and obtain $X/\sigma\sqrt{n} = ULV^T$, where $U$ is a $p \times p$ orthogonal matrix, $U$ is a $n \times n$ orthogonal matrix, and $L \in R^{p \times n}$ with the $(i,i)$-th entry the $i$-th singular value, $\sigma_i$, of $X$. We order the singular values so that $\sigma_1 \geq \sigma_2 \geq \cdots \geq \sigma_{\min(p,n)}$. Third, set $\eta^*(\sigma_i) = \frac{1}{\sigma_i}\sqrt{(\sigma_i^2 - \frac{p}{n} - 1)^2 - 4\frac{p}{n}}$ when $\sigma_i > 1 + \sqrt{\frac{p}{n}}$, or set $\eta^*(\sigma_i) = 0$ otherwise. Finally, the denoised matrix is obtained by $X' := \sigma\sqrt{n}U\tilde{L}V^T$, where $\tilde{L} \in R^{p \times n}$ with the $(i,i)$-th entry $\eta^*(\sigma_i)$. The optimal shrinkage is a nonlinear filtering technique to denoise a noisy matrix, which takes into account the peculiar singular-value/vector structure of $X$ when $p$ and $n$ are on the same scale; that is, when $p = p(n)$ and $\frac{p(n)}{n} \to \gamma > 0$ when $n \to \infty$, the singular vectors and singular values are biased when the data matrix is contaminated by noise. We view our data matrix $\mathcal{X}$ as a composition of the clean signal (the stimulation artifacts), and the noise (the iEEG signal), and then apply the optimal shrinkage to recover the stimulation artifacts. We refer readers with interest in this "large $p$ and large $n$" setup to (Gavish and Donoho, 2017) and citations therein for details. Denote the denoised data matrix as $\mathcal{X}'$, where the i-th column, denoted as $X'_i$, is the denoised artifact cycle of $X_i$.

4. For each $X_i$, determine the $K$ nearest neighbors by the diffusion distance (Coifman and Lafon, 2006; El Karoui and Wu, 2016; Singer and Wu, 2017) determined by $X'_i$, $i = 1, \ldots, n$. Specifically, there are four sub-steps to evaluate the diffusion distance. First, for each $X_i$, find $\mathcal{N}_i := \{X_{i_1}, X_{i_2}, \ldots, X_{i_m}\}$ so that $\|\tilde{X}_{i_1} - \tilde{X}_i\|, \|\tilde{X}_{i_2} - \tilde{X}_i\|, \ldots, \|\tilde{X}_{i_m} - \tilde{X}_i\|$ are minimal, where $m > 0$ is the number chosen by the user; that is, we find all cycles that have the most similar stimulation artifacts. Second, establish a $n \times n$ affinity matrix $W$ so that $W(i, i_j) = exp\left(-\|\tilde{X}_{i_j} - \tilde{X}_i\|^2 / \varepsilon\right)$ when $X_{i_j} \in \mathcal{N}_i$, where $\varepsilon$ is the median of $\left\{\|\tilde{X}_{i_j} - \tilde{X}_i\|\right\}_{\substack{i=1\ldots n \\ j=1\ldots m}}$, and 0 otherwise. With the affinity matrix, establish the degree matrix, a $n \times n$ diagonal matrix $D$ so that the $i$-th diagonal entry is the sum of the $i$-th row of $W$, and hence the $n \times n$ diffusion matrix $A = D^{-1}W$. Note that $A$ can be viewed as a transition matrix of a Markov process defined on the point cloud $\{X_i\}_{i=1}^n$. Since $A$ is similar to a symmetric matrix $D^{-1/2}WD^{-1/2}$ that has the eigendecomposition $V\Lambda V^T$, where $V$ is a $n \times n$ orthogonal matrix and $\Lambda$ is a $n \times n$ diagonal matrix containing eigenvalues $\lambda_1 \geq \lambda_2 \geq \cdots \geq \lambda_n$, we have the decomposition $A = \Phi\Lambda\Psi^T$, where $\Phi = D^{-1/2}V$ and $\Psi = D^{1/2}V$. Third, define $Y_i$ to be the i-th column of $\Lambda^t \Phi^T$, where $t > 0$ is the diffusion time chosen by the user. In general, $Y_i$ is called the diffusion map of $X_i$. The diffusion distance between $X_i$ and $X_j$ is then defined as the Euclidean distance between $Y_i$ and $Y_j$. In this work, we choose $m = 30, t = 1$. Mathematically, the diffusion distance depends on the diffusion process on the point cloud, which incorporates the geometric structure of the point cloud and stabilizes the influence of possible noise (Coifman and Lafon, 2006; El Karoui and Wu, 2016; Singer and Wu, 2017). As a result, the diffusion distance helps us obtain a more accurate cycles that share the same stimulation artifact as that of $X_i$. Denote the $K$ nearest neighbors of $X_i$ associated with the diffusion distance as $X_{i_j}$, $j = 1, \ldots, K$.

5. Apply the Euclidean median on $X_{i_j}$, $j = 1, \ldots, K$, by taking the peak height into account as the weight and obtain the estimated artifact $A'_i$. The Euclidean median is determined by $A'_i :=$ $argmin_{v \in R^p} \sum_{j=1}^{K} w_{i_j} \|X_{i_j} - v\|$ and evaluated via the iteratively reweighted least squares, where the weight $w_{i_j}$ is determined by the peak heights of $X_{i_j}$ and $X_i$ (denoted by $H_{i_j}$ and $H_i$ respectively); that is, $w_{i_j} = \exp\left(-\frac{\|H_i - H_{i_j}\|}{\epsilon}\right)$, where $\epsilon = median\left\{\|H_i - H_{i_j}\|\right\}_{j=1}^{K}$.

6. Remove the artifact $A'_i$ from the recorded iEEG signal and obtain the desired artifact-free iEEG signal. To avoid the boundary effect, the estimated artifact $A'_i$ is tapered by multiplying $A'_i$ by a window function $h \in R^p$ before subtraction, where $h(i) = \sin^2\left(\frac{\pi i}{10}\right)$ when $i = 1, \ldots, 5$, $h(i) = \sin^2\left(\frac{\pi(p-i+1)}{10}\right)$ when $i = 1, \ldots, 5$, and $h(i) = 1$ otherwise.

In this study, we choose $k$ to be 30. Note that we do not intend to remove the noise. $L$ is chosen here to be long enough to cover the possible spiky stimulation artifact. A discussion of the choice of $k$ can be found in the supplementary information.

The time complexity of the SANAR algorithm mainly depends on the SVD for the singular value optimal shrinkage, the nearest neighbor search algorithm for the diffusion distance and nonlocal Euclidean median, and the eigen-decomposition for the diffusion distance. For SVD, since the matrix is full, the time complexity is in general $O(p^2 n)$ when $p \leq n$. For the eigen-decomposition, the complexity is theoretically $O(n^{\omega + \eta})$, where $O(n^\omega)$ is the complexity of the chosen matrix multiplication algorithm, and an arbitrary $\eta > 0$ (Demmel et al., 2007). In general, for the matrix multiplication algorithm, $\omega \approx 2.376$ when the matrix is dense (Coppersmith and Winograd, 1990). For the sparse matrix, $\omega$ can be improved to $2 + \eta'$ for an arbitrary $\eta' > 0$, when the number of neighbors $m$ chosen in the affinity matrix satisfies $m \leq n^{0.14}$ (Yuster and Zwick, 2004). In practice, $m$ usually satisfies $n^{0.14} \leq m \leq n^{0.68}$, so $\omega$ is between $2 + \eta'$ and 2.376. Here we give a conservative bound $O(n^{2.38})$. For the nearest neighbor search algorithm, we count on the k-d tree based algorithm (Friedman et al., 1976), and the time complexity on average is $O((n + K) \log(n))$. We mention that when the dimension $p$ is high, we can consider randomized nearest neighbor search algorithm, like (Jones et al., 2013). The other steps are at most $O(n)$. As a result, the time complex of the proposed SANAR algorithm takes $O(p^2 n + n^{2.38} + (n + K)\log(n))$.

In general, the proposed algorithm might not work if we do not impose any condition. First, although the stimulations are well controlled from time to time, the artifact patterns could vary. However, like other algorithms that aim at removing such artifacts, we presume stimulation to not cause long-lasting changes in brain dynamics (Step 2). Therefore, we could assume that the artifacts $\{A_i\}_{i=1}^n$ could be well parameterized by few parameters; like the height and the width. In a more mathematical terminology, we assume that $\{A_i\}_{i=1}^n$ is identically and independently sampled from a low dimensional manifold. This assumption allows us to apply the diffusion distance to faithfully compare the artifacts in Step 4. Second, we assume that $E_i + N_i$ and $E_j + N_j$ are independent when $i$ is different from $j$, and that $A_i$ and $E_i + N_i$ are jointly independent. This assumption allows us to apply the optimal shrinkage in Step 3 and Euclidean median in Step 5. To sum up, under these two conditions, the proposed algorithm may work.

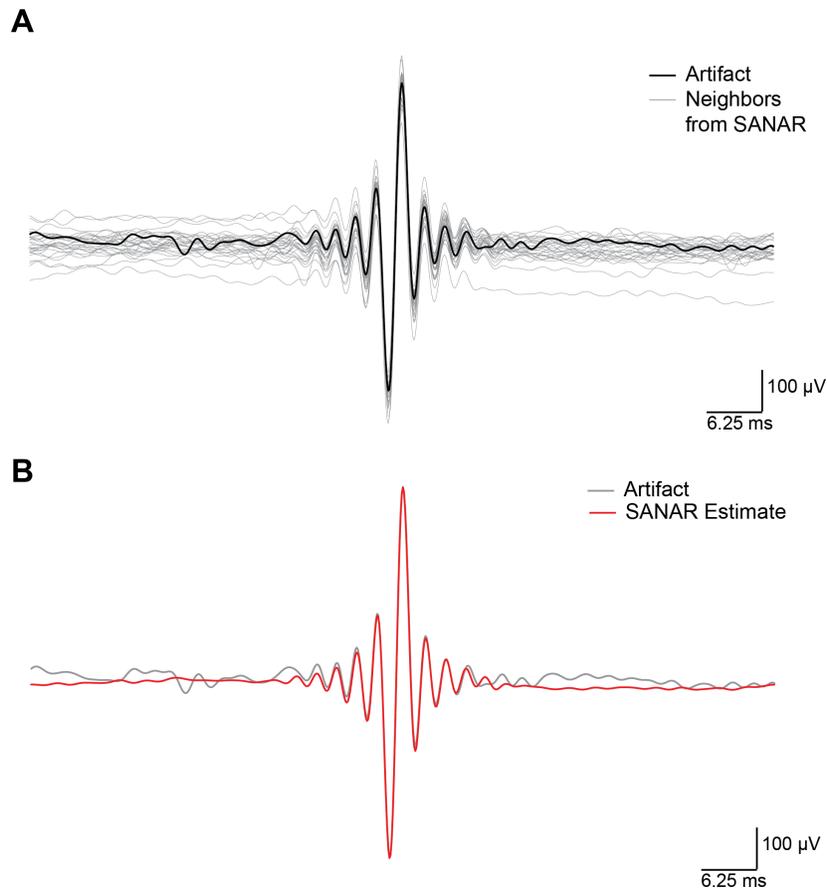

**Figure 2**: Stimulation Artifact Suppression using SANAR. (A) Waveforms of neighbors (gray traces) computed using the diffusion distance method for an example artifact (black trace). (B) Waveforms of artifact template computed from the median of non-local neighbors as used in SANAR (red trace) compared to the artifact waveform (gray trace).

**Stimulation Artifact Removal using Independent Component Analysis.** To remove artifacts using the ICA-based method, we employed a trial-by-trial manual rejection approach (Figure 3). We used the infomax algorithm implemented in EEGlab toolbox for removal of artifacts caused by eye blinks, eye movement and muscle activity (Delorme and Makeig, 2004; Jung et al., 2000) to decompose the iEEG data into independent components. The components that captured the stimulation artifacts were rejected by visual inspection of component waveform and spectra (Figure 3B and 3C). The rest of the components were used to reconstruct the iEEG data free of artifacts. Typically, each trial consisted of 2 components that captured the artifact without containing significant iEEG data ascertained using visual inspection of spectra. In the example shown in Figure 2B and 2C, the components indicated with dashed red box were rejected.

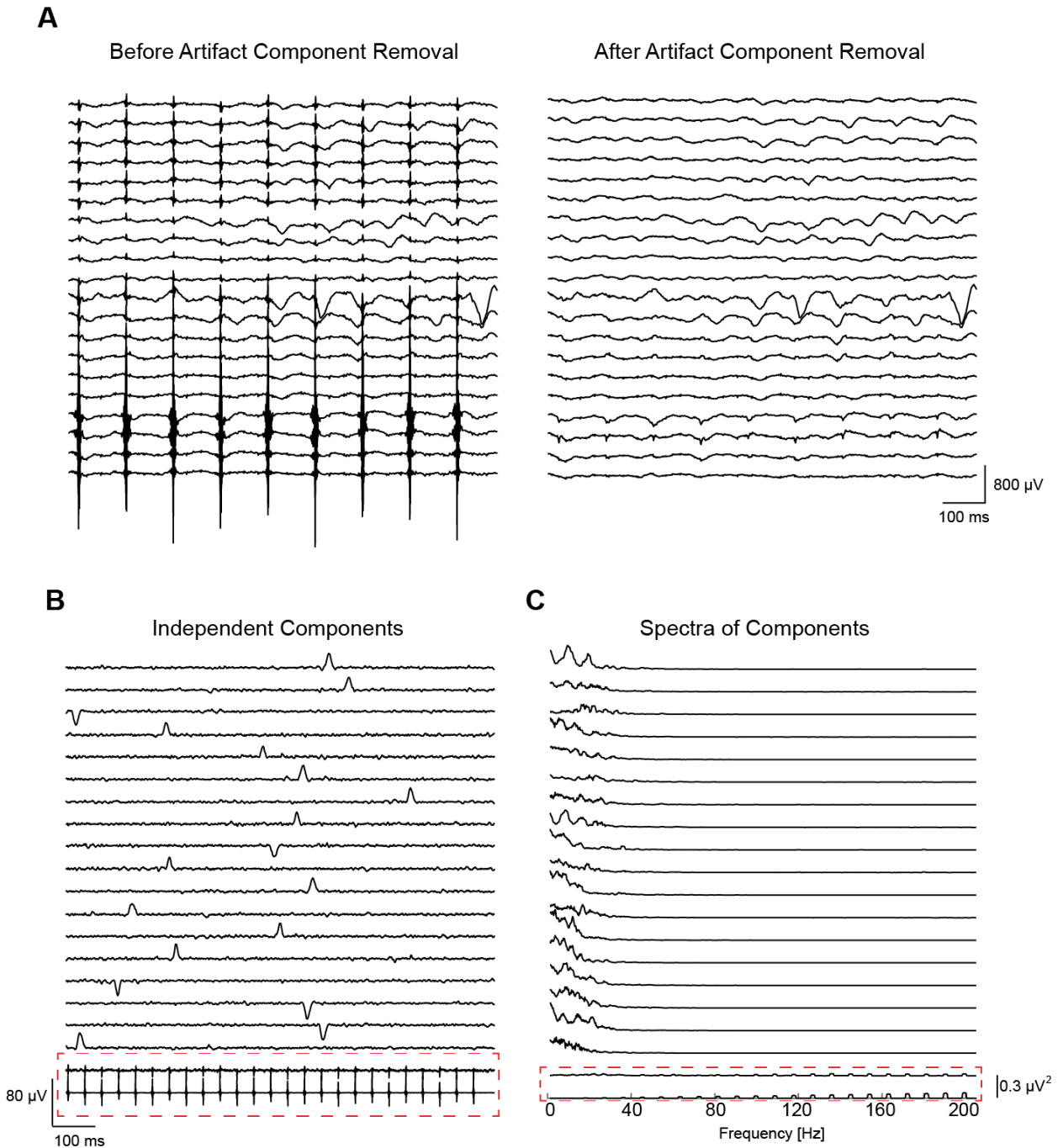

**Figure 3**. Stimulation Artifact Suppression using ICA. (A) Performance of ICA showing effective removal of stimulation artifacts (only 20 of 114 channels shown). (B) Waveform of the independent components obtained by decomposing the data. Red dashed box indicates the components exhibiting stimulation artifact waveform. (C) Spectra of independent components. Red dashed box indicates components exhibiting power in stimulation frequency and harmonics of stimulation frequency.

**Measures of performance.** For the real data, since we do not know the true EEG signal, we consider the following *artifact residue (AR)* index, which was considered in (Malik et al., 2017) for other purposes. For the $i$-th artifact cycle, define the $AR_i$ as

$$AR_i := \left| \log \left( \frac{1}{2} \left[ \frac{med|\tilde{E}_i - med(\tilde{E}_i)|}{med|E_i - med(E_i)|} + \frac{med|E_i - med(E_i)|}{med|\tilde{E}_i - med(\tilde{E}_i)|} \right] \times \frac{1}{2} \left[ \frac{Q95|\tilde{E}_i - med(\tilde{E}_i)|}{max|E_i - med(E_i)|} + \frac{max|E_i - med(E_i)|}{Q95|\tilde{E}_i - med(\tilde{E}_i)|} \right] \right) \right|,$$

where $\tilde{E}_i$ is the estimated iEEG signal recovered over the i-th simulation artifact, $E_i$ is the concatenation of true iEEG signal over the interval without any stimulation artifact from $i - 30, \dots, i + 30$ stimulation artifacts, med means median, Q95 means the 95% quartile, and max means the maximal value. Note that the first term measures how the EEG signal is overall recovered by the algorithm, which is introduced to prevent over-smoothing. The second term measures the presence of artifact residue. Overall, this index captures simultaneously how well the artifact is removed and how well the iEEG signal is recovered. The AR index of a good reconstruction algorithm should be close to 0.

Inspired by the *spectral concentration* (SC) index proposed in (Castells et al., 2005) (Equation (16)), we measure the performance in the frequency domain in the following way. The power spectral density is calculated using Welch's method, featuring 5000 discrete Fourier transform points, Hamming windows of 5000 samples, and 50% overlapping. The 5000 point Fourier transform is chosen to reflect a 5 second window for a frequency resolution of 0.2 Hz. The SC index for the iEEG signal sampled at 1000 Hz is defined as the ratio of the power change (relative to raw signal) over the fundamental frequency and the harmonics of the stimulation frequency to the energy over the rest of the frequencies in the band (1-200 Hz). The band 1-200 Hz is chosen to be sufficiently wide to cover the iEEG spectrum of interest. The canonical frequency band of interest in studies of oscillations extend from 1 to 50 Hz and comprises delta (1 – 4 Hz), theta (4 – 8 Hz), alpha (8 – 13 Hz), beta (13 – 30 Hz) and gamma (30 – 50 Hz) (Fröhlich, 2016). In addition, owing to the direct access to cortical surface in ECoG, activity in the frequency band 70 – 200 Hz also contains information of brain activation. Activity in this band is shown to be correlated with spiking activity (Ray et al., 2008). The SC index of a good reconstruction algorithm should be small.

# Results

**Removal of artifacts from phantom data.** To test the performance of the algorithm in a controlled case where the ground truth is available, we developed a 'phantom' as shown in Figure 4A. In the example shown in Figure 3B, stimulation was applied at 10 Hz and the 'endogenous' signal was a 7 Hz sine wave. The proposed algorithm was effective in removal of the artifact and preserving the spectral content of the 'endogenous' signal. Across all the 12 traces that were cleaned, the AR index was found to be $0.391 \pm 0.018$ while SC was found to be $0.158 \pm 0.016$. Since only 2 traces were available for each trial, it was not possible to apply ICA-based method in this case.

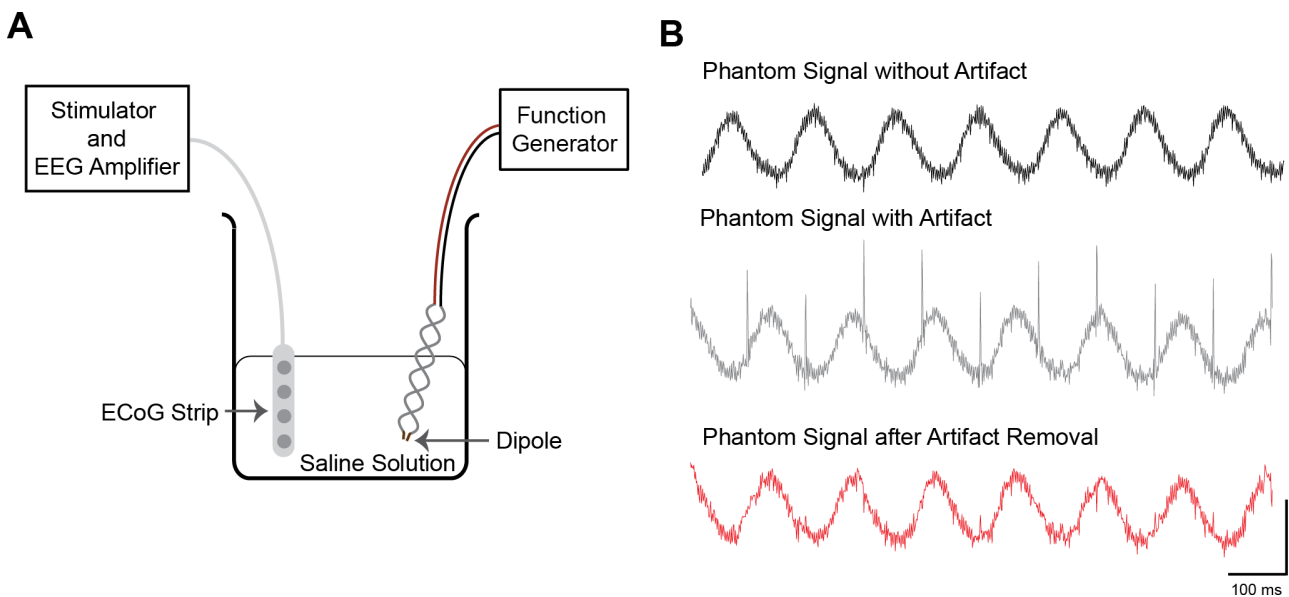

**Figure 4**. Demonstration of SANAR in simulated data. (A) Phantom setup used to simulate stimulation artifacts in the presence of periodic signal. (B) Example traces showing simulated data recorded from ECoG electrodes without artifact, with artifact and after artifact is removed.

**Removal of artifacts from iEEG data.** Both the ICA-based method and our SANAR algorithm appeared to be effective in removing stimulation artifacts when the corresponding waveforms are inspected visually (Fig 5A). However, when the spectra of the signals reconstructed from ICA-based method and SANAR were inspected, it was evident that ICA was effective only in removing the artifact spectral content in the fundamental frequency (stimulation frequency in our case) and first few harmonics of the fundamental frequency (Blue Trace Fig 5B, 5C). In contrast, SANAR was effective in removing artifact spectral content at the fundamental frequency as well as all the harmonics of the fundamental frequency (Red Trace Fig 5B, 5C). Thus, SANAR was more effective in suppressing stimulation artifacts compared to ICA.

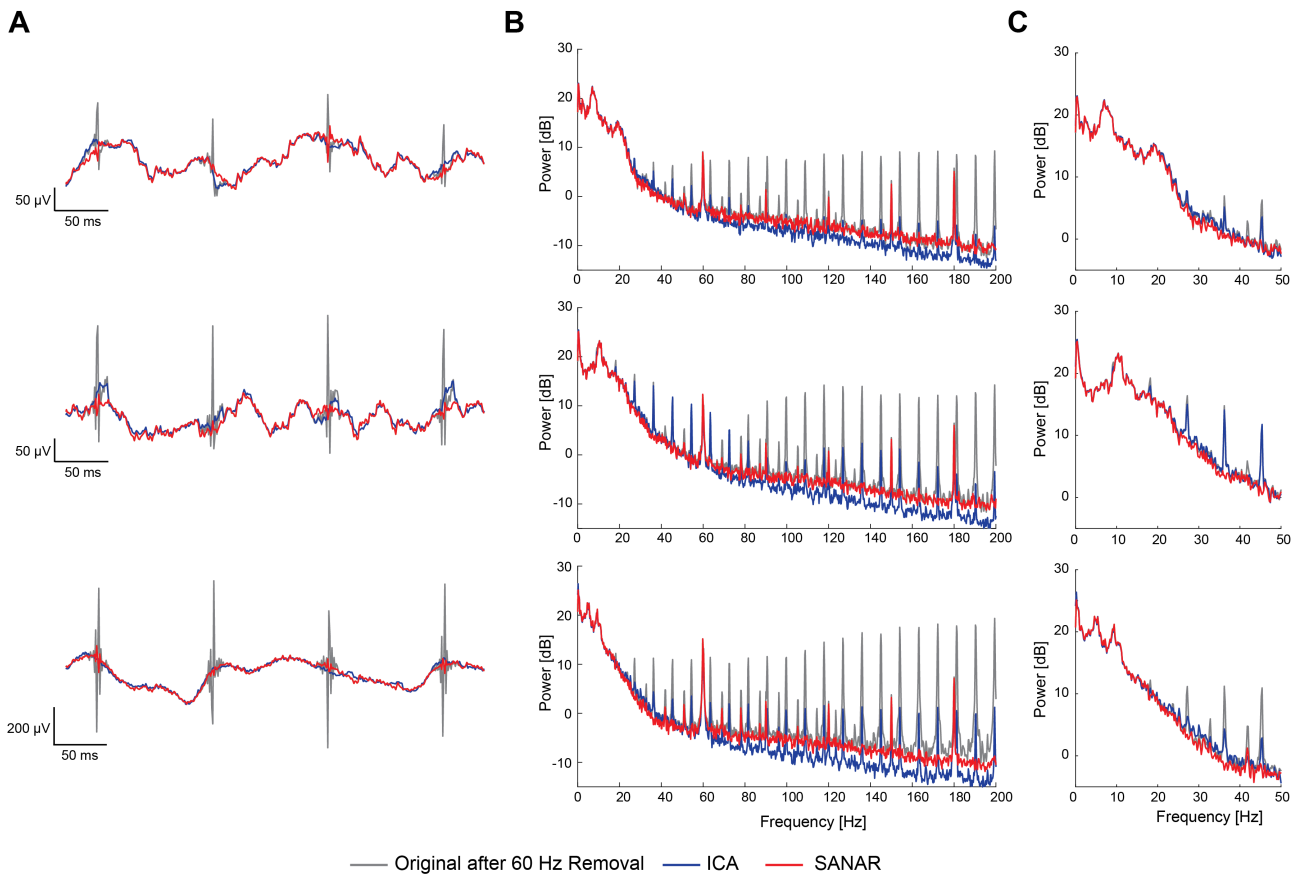

**Figure 5**. Comparison of ICA and SANAR over real signal after removing 60Hz artifact by the sine-wave fitting. (A) Example traces showing the performance of ICA and SANAR in time domain. ICA (Blue trace) tends to produce a smoother interpolation of the data segment in which artifact is present compared to SANAR (Red trace). (B) Spectra corresponding to the channels from which example traces are shown. SANAR is more effective in removing not only the spectral content at stimulation frequency but also the harmonics of the stimulation frequency compared to ICA. (C) Spectra zoomed in to the low frequency region showing that both ICA and SANAR faithfully preserve the spectral content of the endogenous iEEG activity.

Page **17** of **32**

To compare the performance, we computed AR index and SC for the reconstructed signals. The AR index was significantly lower for SANAR (ICA: 0.430 ± 0.015; SANAR: 0.388 ± 0.011 (mean ± s.e.m.) $p < 0.001$ paired t-test), suggesting that SANAR was more effective in removing artifact information in the time domain (Figure 6A). SC was also significantly lower for SANAR (ICA: 0.136 ± 0.002; SANAR: 0.096 ± 0.000 (mean ± s.e.m.) $p < 0.001$ paired t-test) indicating that our proposed algorithm was more effective in suppressing the spectral content of artifacts (Figure 6B). The performance difference of ICA and SANAR with respect to AR index can be considered small (effect size: 0.22, Cohen's d) while the performance difference with respect to SC is quite large (effect size: 2.1, Cohen's d). Moreover, SC was consistently small for all electrodes in case of SANAR compared to ICA while AR was more variable. Thus overall, SANAR's performance was comparable to ICA (albeit slightly better) in the time domain and better than ICA in the frequency domain.

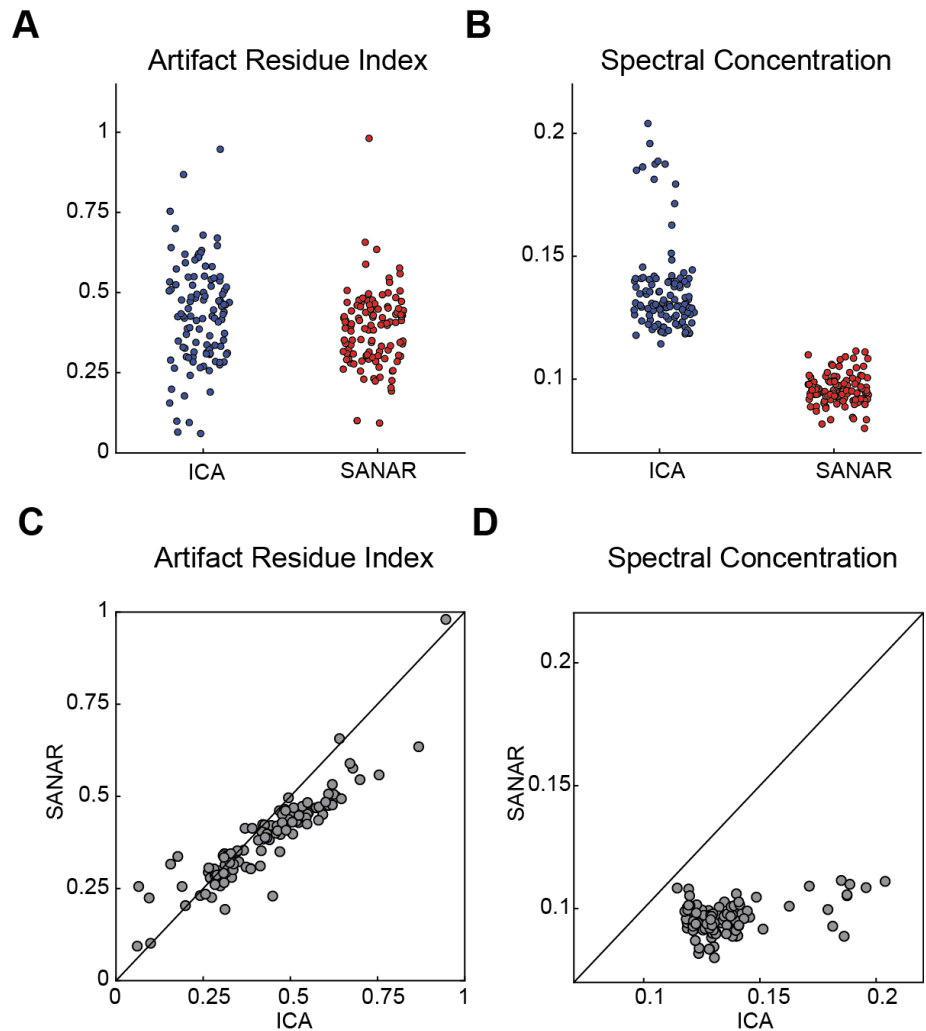

**Figure 6**. Performance comparison of ICA and SANAR. (A) Scatter plot of artifact residue index (AR) computed for ICA and SANAR. Each dot represents an electrode. (B) Scatter plot of spectral concentration (SC) for ICA and SANAR. (C) AR for the two methods for each electrode plotted against each other show that the performance of ICA is better for some electrodes while the performance of SANAR is better for other electrodes. (D) SC for the two methods for each electrode plotted against each other show that performance of SANAR is consistently better than that of ICA

## Discussion and Conclusion

In this work, we have developed an algorithm, SANAR, to effectively remove electrical stimulation artifacts caused by DCS in iEEG data. We have also provided two different metrics that capture the performance of the algorithm quantitatively. The algorithm is able to perform as well as ICA, the current state of the art, in the time domain and exceeds the performance in the frequency domain. This is particularly significant in studies where periodic stimulation is used to study the effect of stimulation on oscillations such as rTMS or DCS. In summary, the ability of SANAR to handle nonstationarity in waveform shape may help overcome the non-linear impact of physiological phenomenon like respiration and heart rate on artifact waveform (Noury et al., 2016). Specifically, the nonstationarity in the stimulation artifacts comes from the variation of brain impedance and other physical quantities induced by the physiological dynamics, like respiration and heart rate. In the case of ICA, these non-stationarities may contribute to less than ideal decomposition into artifacts and EEG components. As our ICA approach has been conservative in rejecting components to preserve as much signal as possible, some residual artifacts are still present at the end of the process.

One of the commonly used approaches to remove stimulation artifacts is the TS method (Alagapan et al., 2016; Hashimoto et al., 2002; Wichmann, 2000) in which artifact waveforms are estimated as a template and subtracted from the neurophysiological data. While the proposed SANAR algorithm can be viewed as a generalization of the TS algorithm, it is essentially different. The traditional TS method is effective when the artifact waveform does not change in shape across trials or time. Due to the nonstationarity, we find templates in a different way; we estimate the artifact pattern directly by an exclusive template for each artifact. The exclusive template does not come from all artifacts or temporally close artifacts, but nonlocally from k-nearest neighbors that have no temporal proximity. Specifically, we respect the nonlinear structure guiding the artifact patterns in this step. Also note that the metric design step for the k-nearest neighbors search does not exist for the TS algorithm. When multiple channels of data are available as in the case of high density EEG or iEEG, PCA or ICA based methods have been used (Lu et al., 2012; Rogasch et al., 2014; ter Braack et al., 2013). The methods involve decomposition of the data contaminated with stimulation artifacts into distinct components that contain artifacts and neurophysiological data. By rejecting components that contain artifacts and reconstruction of remaining components, artifact free data is obtained. These approaches work well when the data is stationary; particularly when the EEG recording is short. However, when the physiological signal is long as is typical in many situations, PCA and ICA may not yield robust decomposition of signal into artifacts. For SANAR, the nonstationarity is fully respected by the "nonlocal" step. Thus, it can perform well even if the signal is long. Also, for the ICA approach, the number of channels needs to be large enough. However, in the proposed algorithm, it can be applied to the single channel signal as seen in the case of the simulated signal. While we have demonstrated the advantages of SANAR compared to ICA in this case report, a more systematic comparison with other possible algorithms, like ensemble Kalman filters and adaptive, is needed to ascertain the advantages of SANAR. Our choice of comparison against ICA was based on the fact that it still widely used in brain stimulation studies (Albouy et al., 2017; Hamidi et al., 2010; Korhonen et al., 2011; Lu et al., 2012).

While the ICA-based method successfully suppressed artifact in the fundamental frequency of the stimulation frequency, the artifact content at higher harmonics were not suppressed sufficiently. This is reflected in the higher SC values that are observed compared to SANAR. The likely explanation is the fact that we removed only those components that did not contain significant spectral content outside the fundamental frequencies and harmonics of stimulation frequency. We regularly found components that had high spectral content in the higher harmonics of the stimulation frequency while also containing spectral content in other frequencies. To preserve the spectral content of the iEEG signals, we did not remove these components. In contrast, SANAR was effective in suppressing the artifact at higher harmonics as the method allows more robust reconstruction of the artifact waveform due to the fact that reconstruction depends on the manifold in which the artifacts exist. Hence, SANAR is effective in removing artifact content in higher frequencies. This property is of specific importance in analysis of iEEG data as the spectral content in higher frequencies reflect spiking activity of the region (Ray et al., 2008) and effective suppression of artifact in these frequency bands are necessary to avoid confounding effect of stimulation.

One significant challenge we faced during the development of SANAR is the interaction between line noise and stimulation artifacts. Since the stimulation artifact was impulse-like, the spectrum at line noise frequency (60 Hz) was confounded with the spectral content of the artifact. Moreover, an amplitude modulation effect was observed with sidebands around 60 Hz (± stimulation frequency). The sine wave fitting approach described in the methods was more effective than a notch filter and resulted in the least distortion in the time domain waveform. In addition, in the frequency domain, the sidebands were significantly reduced using this approach.

While we have demonstrated the algorithm in iEEG data with periodic electrical pulse stimulation, the proposed algorithm has the potential to handle more general artifacts. For example, TMS-induced electrical artifacts that arise due to the interaction between the electric field and the recording equipment (Rogasch et al., 2017; Veniero et al., 2009) in scalp EEG exhibit similar impulse-like characteristics and can be removed with the proposed algorithm albeit with additional modifications that are beyond the scope of the current study. More specifically, TMS also induces a number of additional artifacts like muscle artifacts and sensory evoked artifacts (Hernandez-Pavon et al., 2012; Korhonen et al., 2011; Rogasch et al., 2017; Veniero et al., 2009), and the algorithm proposed here might need to be modified to accommodate these additional artifacts, as the artifact waveform may not be stereotyped. Additionally, in tACS, the stimulation waveform is sinusoidal, and hence the artifact is sine-wave like. Note that while the same model and algorithm could potentially be applied to remove this kind of artifact, the AR index may not be applied since there are no stimulation artifact-free periods. We will explore this direction in the future work.

**Limitations.** While the algorithm provides encouraging results and shows its potential, there are several limitations. One of the main limitations of the currently proposed method is the computation time. In our setup, a desktop with 4-core processor and 32 GB memory running Matlab, we found that SANAR took 54 minutes for running the iEEG data containing 110312 artifacts. As the number of artifacts that must be suppressed increase, the computational time required to identify neighbors based on the manifold also increases. Since the designed metric is not Euclidean, we cannot count on the existing nearest neighbor search algorithms to find nearest neighbors efficiently. We thus need to develop a nearest neighbor search algorithm for the des metric. We could also use available surrogate information to narrow down the possible neighbor candidates; for example, if heights of two artifacts are very different, we do not expect them to be neighbors, and we can focus on finding neighbors from those beats with similar heights. A more systematic approach to treat the computational issue is needed to handle the high-throughput data in the near future. The processing time of the infomax algorithm used in the study to decompose one trial (3500 samples) of 114 channels into independent components was approximately 15 seconds on average. Thus, for the dataset used in the study with 33 trials, the computation time was approximately 8 minutes. However, each ICA computation was interleaved with manual rejection of components by visual inspection of component waveform and component spectra which was variable between the trials. Therefore, the entire process of rejection took approximately 30 minutes.

Another limitation of SANAR is the knowledge of stimulation times. This is not a big issue if we could determine those artifacts relatively easily throughout the dataset, for example, when there is a clear landmark, like in the examples in this paper. If the artifact has morphology without a clear landmark, or when there are multiple stimulations with different morphologies, SANAR cannot be applied directly. We need to combine other techniques with SANAR to handle the signal. For example, when there is a regular pattern of the appearance of artifacts, the recently developed de-shape short time Fourier transform could help to determine the time stamps of artifacts, despite the artifact morphology. This kind of approach has been developed for the maternal abdominal electrocardiogram signal to extract the fetal electrocardiogram signal. Although we expect a similar approach to handle this limitation, a systematic exploration is needed to confirm the performance.

From the algorithmic viewpoint, there are few parts that could be improved, pending the theoretical development. For example, what is the optimal number of nearest neighbors for the algorithm? What is the best shrinkage policy when we remove the EEG signal for the metric design? What is the optimal metric when we determine the neighbors? While the chosen parameters and designed metric work efficiently, we expect to improve the performance by taking statistical development into account. The above-mentioned algorithmic and theoretical challenges will be explored in future work.

## Funding

Research reported in this publication was supported in part by the National Institute of Mental Health of the National Institutes of Health under Award Numbers R01MH101547 and R21MH105557, National Institute of Neurological Disorders and Stroke of the National Institutes of Health under award number R21NS094988-01A1. The content is solely the responsibility of the authors and does not necessarily represent the official views of the National Institutes of Health.

# References

1. Al-ani T, Cazettes F, Palfi S, Lefaucheur JP. Automatic removal of high-amplitude stimulus artefact from neuronal signal recorded in the subthalamic nucleus. Journal of neuroscience methods, 2011; 198: 135-46.
2. Alagapan S, Schmidt SL, Lefebvre J, Hadar E, Shin HW, Frhlich F. Modulation of Cortical Oscillations by Low-Frequency Direct Cortical Stimulation Is State-Dependent. PLoS biology, 2016; 14: e1002424.
3. Albouy P, Weiss A, Baillet S, Zatorre RJ. Selective Entrainment of Theta Oscillations in the Dorsal Stream Causally Enhances Auditory Working Memory Performance. Neuron, 2017.
4. Allen DP, Stegemoller EL, Zadikoff C, Rosenow JM, Mackinnon CD. Suppression of deep brain stimulation artifacts from the electroencephalogram by frequency-domain Hampel filtering. Clinical neurophysiology : official journal of the International Federation of Clinical Neurophysiology, 2010; 121: 1227-32.
5. Borchers S, Himmelbach M, Logothetis N, Karnath HO. Direct electrical stimulation of human cortex - the gold standard for mapping brain functions? Nature reviews. Neuroscience, 2012; 13: 63-70.
6. Castells F, Rieta JJ, Millet J, Zarzoso V, Associate. Spatiotemporal blind source separation approach to atrial activity estimation in atrial tachyarrhythmias. IEEE transactions on bio-medical engineering, 2005; 52: 258-67.
7. Chaudhury KN, Singer A. Non-Local Euclidean Medians. IEEE Signal Process Lett, 2012; 19: 745-8.
8. Coifman RR, Lafon S. Diffusion maps. Appl Comput Harmon A, 2006; 21: 5-30.
9. Coppersmith D, Winograd S. Matrix Multiplication Via Arithmetic Progressions. J Symb Comput, 1990; 9: 251-80.
10. Delorme A, Makeig S. EEGLAB: an open source toolbox for analysis of single-trial EEG dynamics including independent component analysis. Journal of neuroscience methods, 2004; 134: 9-21.
11. Demmel J, Dumitriu I, Holtz O. Fast linear algebra is stable. Numer Math, 2007; 108: 59-91.
12. Duffau H, Capelle L, Sichez J, Faillot T, Abdennour L, Law Koune JD, Dadoun S, Bitar A, Arthuis F, Van Effenterre R, Fohanno D. Intra-operative direct electrical stimulations of the central nervous system: the Salpetriere experience with 60 patients. Acta neurochirurgica, 1999; 141: 1157-67.

13. El Karoui N, Wu HT. Graph Connection Laplacian Methods Can Be Made Robust to Noise. Ann Stat, 2016; 44: 346-72.
14. Ezzyat Y, Kragel JE, Burke JF, Levy DF, Lyalenko A, Wanda P, O'Sullivan L, Hurley KB, Busygin S, Pedisich I, Sperling MR, Worrell GA, Kucewicz MT, Davis KA, Lucas TH, Inman CS, Lega BC, Jobst BC, Sheth SA, Zaghloul K, Jutras MJ, Stein JM, Das SR, Gorniak R, Rizzuto DS, Kahana MJ. Direct Brain Stimulation Modulates Encoding States and Memory Performance in Humans. Current biology : CB, 2017; 27: 1251-8.
15. Ezzyat Y, Wanda PA, Levy DF, Kadel A, Aka A, Pedisich I, Sperling MR, Sharan AD, Lega BC, Burks A, Gross RE, Inman CS, Jobst BC, Gorenstein MA, Davis KA, Worrell GA, Kucewicz MT, Stein JM, Gorniak R, Das SR, Rizzuto DS, Kahana MJ. Closed-loop stimulation of temporal cortex rescues functional networks and improves memory. Nature communications, 2018; 9: 365.
16. Friedman JH, Bentley JL, Finkel RA. An algorithm for finding best matches in logarithmic time. ACM Trans. Math. Software, 1976; 3: 209-26.
17. Frohlich F. Endogenous and exogenous electric fields as modifiers of brain activity: rational design of noninvasive brain stimulation with transcranial alternating current stimulation. Dialogues in clinical neuroscience, 2014; 16: 93-102.
18. Fröhlich F. Network neuroscience. Academic Press: Amsterdam ; Boston, 2016.
19. Gavish M, Donoho DL. Optimal Shrinkage of Singular Values. Ieee T Inform Theory, 2017; 63: 2137-52.
20. Hamidi M, Slagter HA, Tononi G, Postle BR. Brain responses evoked by high-frequency repetitive transcranial magnetic stimulation: an event-related potential study. Brain stimulation, 2010; 3: 2-14.
21. Hashimoto T, Elder CM, Vitek JL. A template subtraction method for stimulus artifact removal in high-frequency deep brain stimulation. Journal of neuroscience methods, 2002; 113: 181-6.
22. Helfrich RF, Schneider TR, Rach S, Trautmann-Lengsfeld SA, Engel AK, Herrmann CS. Entrainment of brain oscillations by transcranial alternating current stimulation. Current biology : CB, 2014; 24: 333-9.
23. Hernandez-Pavon JC, Metsomaa J, Mutanen T, Stenroos M, Maki H, Ilmoniemi RJ, Sarvas J. Uncovering neural independent components from highly artifactual TMS-evoked EEG data. Journal of neuroscience methods, 2012; 209: 144-57.
24. Jones PW, Osipov A, Rokhlin V. A randomized approximate nearest neighbors algorithm.

Appl Comput Harmon A, 2013; 34: 415-44.
25. Jung TP, Makeig S, Humphries C, Lee TW, McKeown MJ, Iragui V, Sejnowski TJ. Removing electroencephalographic artifacts by blind source separation. Psychophysiology, 2000; 37: 163-78.
26. Kocabicak E, Temel Y, Hollig A, Falkenburger B, Tan SKH. Current perspectives on deep brain stimulation for severe neurological and psychiatric disorders. Neuropsychiatric disease and treatment, 2015; 11: 1051-66.
27. Korhonen RJ, Hernandez-Pavon JC, Metsomaa J, Maki H, Ilmoniemi RJ, Sarvas J. Removal of large muscle artifacts from transcranial magnetic stimulation-evoked EEG by independent component analysis. Med Biol Eng Comput, 2011; 49: 397-407.
28. Kucewicz MT, Berry BM, Miller LR, Khadjevand F, Ezzyat Y, Stein JM, Kremen V, Brinkmann BH, Wanda P, Sperling MR, Gorniak R, Davis KA, Jobst BC, Gross RE, Lega B, Van Gompel J, Stead SM, Rizzuto DS, Kahana MJ, Worrell GA. Evidence for verbal memory enhancement with electrical brain stimulation in the lateral temporal cortex. Brain : a journal of neurology, 2018.
29. Laguna P, Sornmo L. Sampling rate and the estimation of ensemble variability for repetitive signals. Med Biol Eng Comput, 2000; 38: 540-6.
30. Lu YL, Cao PJ, Sun JJ, Wang J, Li LM, Ren QS, Chen Y, Chai XY. Using independent component analysis to remove artifacts in visual cortex responses elicited by electrical stimulation of the optic nerve. Journal of neural engineering, 2012; 9.
31. Luber B, Lisanby SH. Enhancement of human cognitive performance using transcranial magnetic stimulation (TMS). NeuroImage, 2014; 85: 961-70.
32. Luo S, Johnston P. A review of electrocardiogram filtering. Journal of electrocardiology, 2010; 43: 486-96.
33. Malik J, Reed N, Wang CL, Wu HT. Single-lead f-wave extraction using diffusion geometry. Physiol Meas, 2017; 38: 1310-34.
34. Matsumoto R, Nair DR, LaPresto E, Najm I, Bingaman W, Shibasaki H, Luders HO. Functional connectivity in the human language system: a cortico-cortical evoked potential study. Brain : a journal of neurology, 2004; 127: 2316-30.
35. Mitra PP, Pesaran B. Analysis of dynamic brain imaging data. Biophys J, 1999; 76: 691-708.
36. Morbidi F, Garulli A, Prattichizzo D, Rizzo C, Manganotti P, Rossi S. Off-line removal of TMS-induced artifacts on human electroencephalography by Kalman filter. Journal of neuroscience methods, 2007; 162: 293-302.

37. Morbidi F, Garulli A, Prattichizzo D, Rizzo C, Rossi S. Application of Kalman Filter to Remove TMS-Induced Artifacts from EEG Recordings. Ieee T Contr Syst T, 2008; 16: 1360-6.
38. Noury N, Hipp JF, Siegel M. Physiological processes non-linearly affect electrophysiological recordings during transcranial electric stimulation. NeuroImage, 2016.
39. Qian X, Chen Y, Feng Y, Ma B, Hao H, Li L. A Method for Removal of Deep Brain Stimulation Artifact From Local Field Potentials. IEEE transactions on neural systems and rehabilitation engineering : a publication of the IEEE Engineering in Medicine and Biology Society, 2017; 25: 2217-26.
40. Ray S, Crone NE, Niebur E, Franaszczuk PJ, Hsiao SS. Neural correlates of high-gamma oscillations (60-200 Hz) in macaque local field potentials and their potential implications in electrocorticography. J Neurosci, 2008; 28: 11526-36.
41. Rogasch NC, Sullivan C, Thomson RH, Rose NS, Bailey NW, Fitzgerald PB, Farzan F, Hernandez-Pavon JC. Analysing concurrent transcranial magnetic stimulation and electroencephalographic data: A review and introduction to the open-source TESA software. NeuroImage, 2017; 147: 934-51.
42. Rogasch NC, Thomson RH, Farzan F, Fitzgibbon BM, Bailey NW, Hernandez-Pavon JC, Daskalakis ZJ, Fitzgerald PB. Removing artefacts from TMS-EEG recordings using independent component analysis: importance for assessing prefrontal and motor cortex network properties. NeuroImage, 2014; 101: 425-39.
43. Santillan-Guzman A, Heute U, Muthuraman M, Stephani U, Galka A. DBS Artifact Suppression using a Time-Frequency Domain Filter. 2013 35th Annual International Conference of the Ieee Engineering in Medicine and Biology Society (Embc), 2013: 4815-8.
44. Singer A, Wu HT. Spectral convergence of the connection Laplacian from random samples. Inf Inference, 2017; 6: 58-123.
45. Sun Y, Farzan F, Garcia Dominguez L, Barr MS, Giacobbe P, Lozano AM, Wong W, Daskalakis ZJ. A novel method for removal of deep brain stimulation artifact from electroencephalography. Journal of neuroscience methods, 2014; 237: 33-40.
46. ter Braack EM, de Jonge B, van Putten MJAM. Reduction of TMS Induced Artifacts in EEG Using Principal Component Analysis. Ieee T Neur Sys Reh, 2013; 21: 376-82.
47. Trebaul L, Rudrauf D, Job AS, Maliia MD, Popa I, Barborica A, Minotti L, Mindruta I, Kahane P, David O. Stimulation artifact correction method for estimation of early cortico-cortical evoked potentials. Journal of neuroscience methods, 2016; 264: 94-102.

48. Veniero D, Bortoletto M, Miniussi C. TMS-EEG co-registration: on TMS-induced artifact. Clinical neurophysiology : official journal of the International Federation of Clinical Neurophysiology, 2009; 120: 1392-9.
49. Wichmann T. A digital averaging method for removal of stimulus artifacts in neurophysiologic experiments. Journal of neuroscience methods, 2000; 98: 57-62.
50. Yuster R, Zwick U. Fast sparse matrix multiplication. Lect Notes Comput Sc, 2004; 3221: 604-15.

# Supplementary Information

**Effect of k on performance**

To estimate the effect of the number of nearest neighbors "k" has on the performance of SANAR, we ran the algorithm on real data and simulated data with different values of k. See Figure S1 for the result. The mean of all realizations is plotted as the solid curve, and the mean +/- standard deviation are plotted as the dashed curves. The red dash lines are provided to enhance the visualization. We varied k from 5 to 50 (in increments of 1 for simulated data and increments of 5 for the real data). In both the real and simulated datasets, increasing k resulted in a decrease in spectral concentration and an increase in artifact residual index. This phenomenon comes from the fact that SANAR is a time domain-based algorithm. The fewer the neighboring stimulation artifacts we choose, the more similar these stimulation artifacts are. Thus, the portion of stimulation artifact after taking median is less deformed, but with a larger variation. As a consequence, the EEG recovery in the frequency domain is less ideal when k is small. This discrepancy could be viewed as a trade-off between time-domain and frequency-domain information recovery. Although we do not optimize k for the algorithm but take the commonly chosen value in diffusion geometry society, based on figures below, the choice of k=30 leads to a balanced EEG recovery in the time and spectral domains. For a specific application, it might be beneficial to fine-tune k to achieve a better performance.

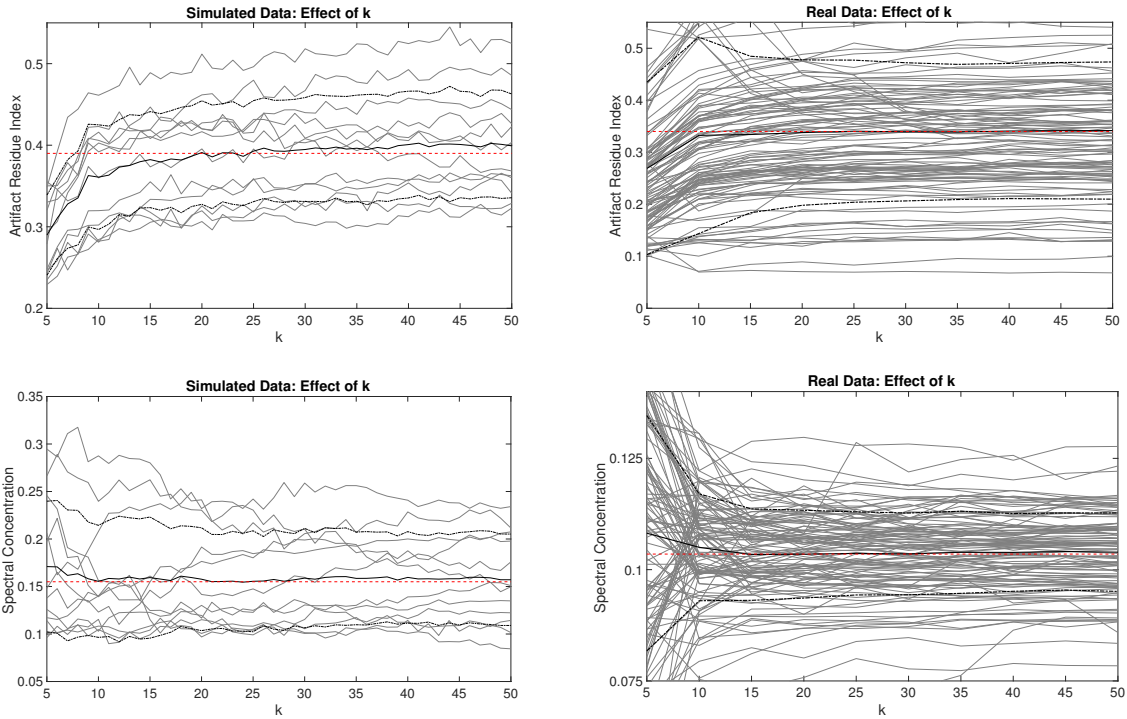

**Figure S1**: The impact of k, the number of nearest neighbors, on the performance of SANAR. The mean of all realizations is plotted as the solid curve, and the mean +/- standard deviation are plotted as the dashed curves. The red dash lines are provided to enhance the visualization.

**Effect of Noise on Performance**

While the simulated data we used in our study was an idealized condition in terms of signal, the stimulation artifacts and the noise in the data were close to real-world conditions. We used the exact set-up as is used in the actual intracranial stimulation experiments and the conductivity of the saline solution is similar to the conductivity found in the brain resulting in artifacts and noise that mimic the actual data.

Before exploring the level of "noise", we would like to make clear the terminologies. In this work, we do not distinguish between EEG and noise, or aim to denoise the EEG signal. We view EEG and noise together as "noise" when we estimate the stimulation artifact. Now we provide the effect of the level of noise in the following way. In our implementation of the algorithm, the artifact is the signal of interest as we are trying to get the best possible estimate of the artifact. So, one way of quantifying the effect of noise would be to look at the performance vs amplitude of artifact. High amplitude artifacts would correspond to high signal-to-noise ratio while low amplitude artifacts would correspond to low signal-to-noise ratio. Since our real-data already has a good combination of these different scenarios, we ran a correlation between the performance measures and stimulation artifact amplitudes (before running our algorithm). We used spearman's rho as the distribution was not normal. We found that there is a strong negative correlation between stimulation artifact amplitude and artifact residue index while there is no correlation between stimulation artifact amplitude and spectral concentration. The results imply as the signal-to-noise ratio increases, the time-domain performance of the algorithm increases while the frequency domain performance of the algorithm remains relatively unchanged. This is not entirely surprising as the algorithm is essentially a time-domain algorithm.

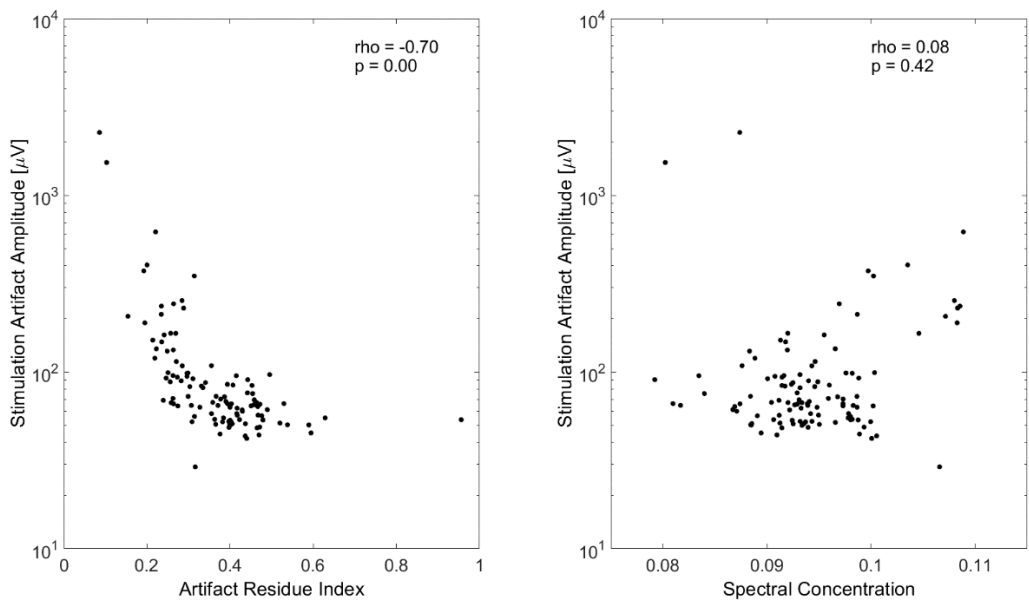

**Figure S2.** Effect of stimulation artifact amplitude on SANAR performance. The AR index was lower for higher amplitude artifacts while SC was unaffected by artifact amplitudes.

**60Hz noise removal**

We provide more information about 60Hz noise removal in SANAR algorithm. The graphs presented on Figure 4 are after 60 Hz removal using the curve fitting approach. The method while effective is not perfect. Figure S3 shows an example electrode which demonstrates the performance of the sine-wave fitting method. Note that we can apply the notch filter after recovering the EEG signal, but we do not do it for the sake of showing readers the effect of the whole algorithm.

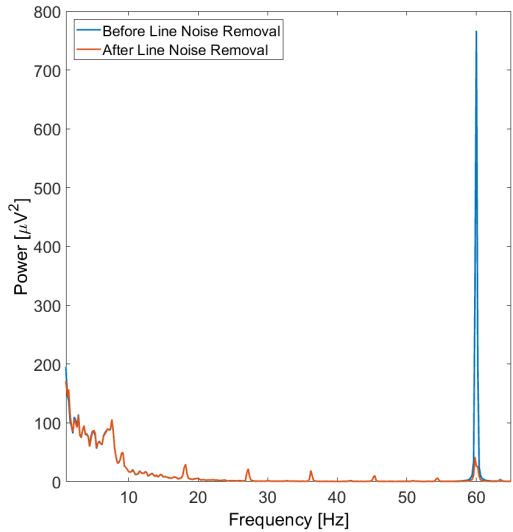

**Figure S3**. The effect of the sine-wave fitting.

The performance of the chosen sine-wave fitting approach seems poorer than a notch filter in terms of attenuation. However, notch filters may introduce distortions (Mitra & Pesaran, 1999). Moreover, we observed that notch filtering before running SANAR resulted in worse performance compared to sine wave fitting. See Figure S4 for a result comparing the notch filter and the sine-wave fitting. Thus, in the context of our algorithm, the sine-wave fitting approach is more advantageous.

Next, we show the amplitude modulation effect caused by the interaction between line noise and stimulation artifacts. Before that, we mention that the amplifier in our experiment has a 500 Hz anti-alias filter that would prevent aliasing. For the amplitude modulation, in the case of pure-sinusoidal stimulation artifact, amplitude modulation results in side band peaks at $F_c + F_m$ and $F_c - F_m$, where $F_c$ is the frequency of the line noise and $F_m$ is the frequency of stimulation artifact. However, when the stimulation artifact is non-sinusoidal, additional sideband peaks occur at $F_c + n*F_m$ and $F_c - n*F_m$, where $n*F_m$ refers to the n-th harmonic of the frequency of the stimulation artifact. In the spectra shown in Figure S4, we see peaks associated with these sideband frequencies in the spectra. (For the sake of clarity only the first 5 harmonics are denoted by lines). In Figure S5, the stimulation frequency is $F_m = 9.09$ Hz, line noise frequency $F_c = 60$ Hz. Additional investigation is required to identify the physiological cause of this amplitude modulation.

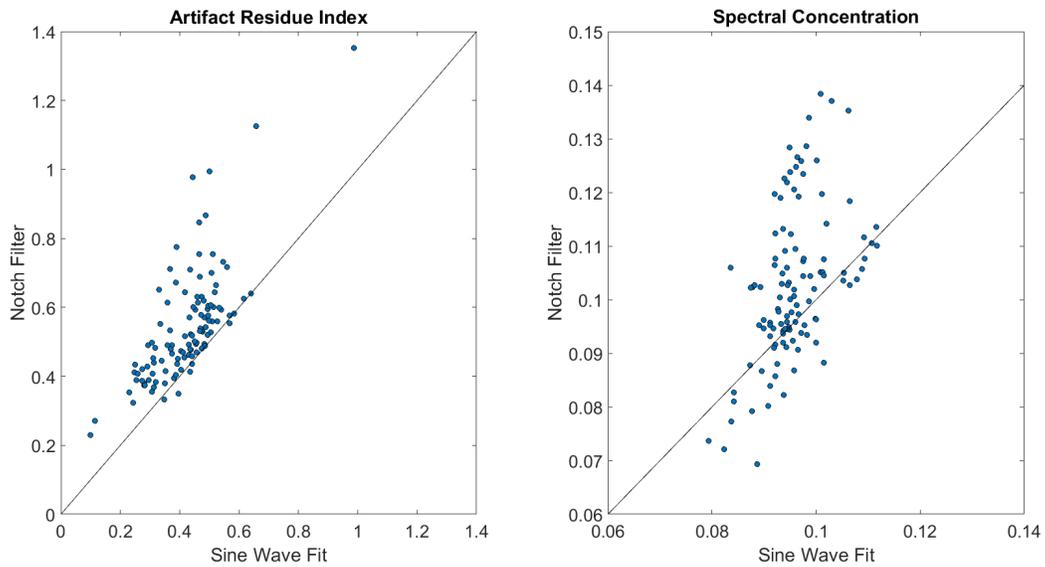

**Figure S4**. Comparison of sine wave fitting and notch filtering before SANAR in terms of performance. Each dot represents an electrode from the ECoG dataset.

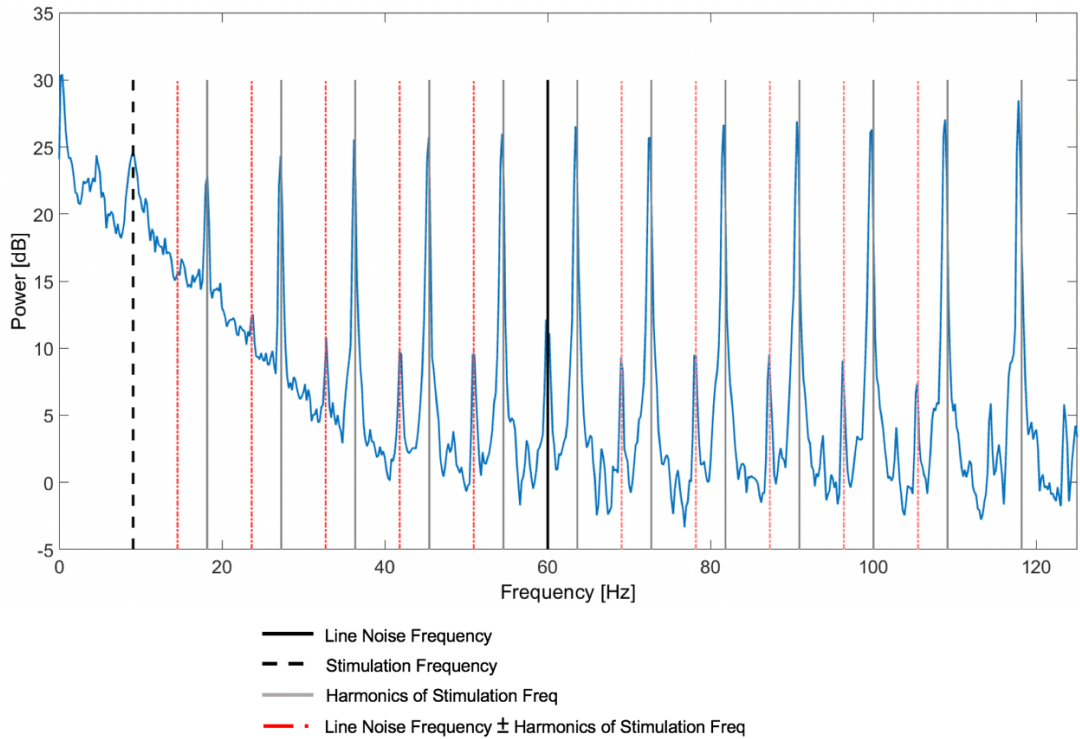

**Figure S5**. the amplitude modulation effect caused by the interaction between line noise and stimulation artifacts.